\documentclass[12pt]{article}
\usepackage{graphicx}

\newcommand\pubnumber{}
\newcommand\pubdate{}

\def\Title#1{\begin{center} {\Large #1 } \end{center}}
\def\Author#1{\begin{center}{ \sc #1} \end{center}}
\def\Address#1{\begin{center}{ \it #1} \end{center}}

\newcommand\pubblock{\rightline{\begin{tabular}{l} \pubnumber\\
         \pubdate  \end{tabular}}}
\newenvironment{Abstract}{\begin{quotation}  }{\end{quotation}}
\newenvironment{Presented}{\begin{quotation} \begin{center} 
             PRESENTED AT\end{center}\bigskip 
      \begin{center}\begin{large}}{\end{large}\end{center} \end{quotation}}

\def\beq{\begin{equation}}
\def\eeq#1{\label{#1}\end{equation}}
\def\eeqn{\end{equation}}

\def\beqa{\begin{eqnarray}}
\def\eeqa#1{\label{#1}\end{eqnarray}}
\def\eeqan{\end{eqnarray}}

\let\bar=\overbar

\def\Dslash{\not{\hbox{\kern-4pt $D$}}}
\def\dslash{\not{\hbox{\kern-2pt $\del$}}}

\def\msb{{\bar{\ssstyle M \kern -1pt S}}}

\begin{document}
\begin{titlepage}
\pubblock

\vfill
\Title{Theoretical uncertainty in $\sin2\beta$: An update}
\vfill
\Author{Marco Ciuchini$\,^{a,}$\footnote{Talk given by M.C.}, Maurizio Pierini$\,^b$,
and Luca Silvestrini$\,^c$}
\Address{$^a$ INFN Sezione di Roma Tre, Via della Vasca Navale 84, 00146 Roma, Italy\\
$^b$ CERN, CH-1211, Geneva 23, Switzerland\\
$^c$ INFN Sezione di Roma, Piazzale Aldo Moro 2, 00185 Roma, Italy}
\vfill
\begin{Abstract}
The source of theoretical uncertainty in the extraction of $\sin2\beta$ from the 
measurement of the golden channel $B_d\to J/\psi K^0$ is briefly reviewed. An updated
estimate of this uncertainty based on $SU(3)$ flavour symmetry and the measurement of
the decay $B_d\to J/\psi\pi^0$ is also presented.
\end{Abstract}
\vfill
\begin{Presented}
Proceedings of CKM2010, the 6th International Workshop on the CKM Unitarity Triangle,
University of Warwick, UK, 6-10 September 2010
\end{Presented}
\vfill
\end{titlepage}
\def\thefootnote{\fnsymbol{footnote}}
\setcounter{footnote}{0}

\section{Introduction}
\label{sec:intro}

The decay $B_d\to J/\psi K$ was recognized long ago as a golden mode for extracting
$\sin2\beta$, $\beta$ being one of the angles of the Unitarity Triangle (UT), by
measuring the time-dependent CP asymmetry~\cite{Bigi:1981qs}.
In fact, its decay amplitude is strongly dominated by a single term with the consequence
that the hadronic uncertainties largely cancel out in the CP asymmetry, making this
measurement the prototype of the ``theoretically clean'' measurements in $B$
physics.

Yet a subleading amplitude with a different weak phase, however small, is present and
introduces a theoretical uncertainty in the extraction of $\sin2\beta$. This uncertainty
needs to be evaluated, in view of the remarkable accuracy on the measurement of the
$B_d \to J/\psi K$ CP asymmetry reached at the $B$ factories and even more of the high
precision expected at LHCb and at the next-generation super $B$ factories.

Unfortunately, no reliable purely theoretical estimate of the $B_d\to J/\psi K$ decay
amplitude is available as this amplitude does not factorize and is not readily computable
using non-perturbative techniques, such as lattice QCD or QCD sum rules.
However, back in 1999, Robert Fleischer pointed out that the theoretical error in the extraction
of $\sin2\beta$ from the $B_d\to J/\psi K$ CP asymmetry could be estimated from data
using the decay $B_s\to J/\psi K_{S,L}$ and the $SU(3)$ flavour symmetry with no additional
assumptions~\cite{hep-ph/9903455}. At that time, however, no measurement of
$B_s\to J/\psi K_{S,L}$ was available. Recently CDF measured the CP-averaged Branching
Ratio ($BR$)~\cite{CDF10240}, but the time-dependent analysis is still missing. For this reason
the method cannot be used yet.

Later on, we proposed an alternative method still based on flavour
symmetry, but requiring few additional assumptions on hadronic
amplitudes~\cite{hep-ph/0507290}.
This method makes use of time-dependent measurements of the channel $B_d\to
J/\psi \pi^0$ available from the $B$ factories. We obtained an
estimate of the theoretical uncertainty in the extraction of $\sin2\beta$
that was non-negligible with respect to the experimental errors. A conservative
evaluation of $SU(3)$-breaking effects was used in the absence of
additional experimental information.

More recently, an updated analysis based on this method
appeared. Using more precise data and an estimate
of $SU(3)$ breaking mainly based on factorization, the authors of 
ref.~\cite{arXiv:0809.0842} found that $\beta$ could be shifted by
as much as $[-3.9,-0.8]^\circ$ at $1\sigma$.

In this proceedings, we briefly review the issue of theoretical uncertainties
in the extraction of $\sin2\beta$ from $B_d \to J/\psi K^0$ and present
an update of our estimate based on ref.~\cite{hep-ph/0507290}.

\section{CP violation in $\mathbf{B_d \to J/\psi K^0}$}
\label{sec:sin2b}

The decay amplitude for $B_d \to J/\psi K^0$ in the Standard Model
(SM) can be written as
\begin{eqnarray}
  \mathcal{A}(B_d \to J/\psi K^0) &=&
  \lambda_c^s A_c(B_d,K^0) - \lambda_u^s A_u(B_d,K^0) \nonumber \\
  \mathcal{A}(\bar B_d \to J/\psi \bar
  K^0) &=& \lambda_c^{s*} A_c(B_d,K^0) - 
  \lambda_u^{s*} A_u(B_d,K^0)\,,                   
  \label{eq:ABdK0}
\end{eqnarray}
where $\lambda_{u_i}^{d_j}=V_{u_id_j}V^*_{u_ib}$ and $V$ is the
Cabibbo-Kobayashi-Maskawa (CKM) matrix
\cite{Cabibbo:1963yz,Kobayashi:1973fv}. The hadronic amplitudes
$A_{u,c}(B_d,K^0)$ can be written in terms of the Renormalization Group
Invariant (RGI) amplitudes introduced in ref.~\cite{hep-ph/9812392} as
\begin{eqnarray}
  A_c(B_d,K^0) &=& E_2(c,c,s;B_d,J/\psi,K^0) +
  P_2(s,c;B_d,J/\psi,K^0)\,,\nonumber \\
  A_u(B_d,K^0) &=& P_2^\mathrm{GIM}(s,c;B_d,J/\psi,K^0) -
  P_2(s,c;B_d,J/\psi,K^0)\,.
  \label{eq:hadparBdK0}
\end{eqnarray}

The time-dependent CP asymmetry in $B_d \to J/\psi K_{S,L}$ is given by
\begin{eqnarray}
  \label{eq:acp}
    a_{CP}^{B_d \to J/\psi K_{S,L}}(t)&=&\frac{\Gamma(B_d(t)\to
    J/\psi K_{S,L})-\Gamma({\bar B}_d(t)\to
    J/\psi K_{S,L})}{\Gamma(B_d(t)\to
    J/\psi K_{S,L})+\Gamma({\bar B}_d(t)\to
    J/\psi K_{S,L})}=\nonumber\\
    && C_{B_d \to J/\psi K_{S,L}}
  \cos(\Delta m_{B_d} t) - S_{B_d\to J/\psi K_{S,L}} \sin(\Delta m_{B_d} t)\,,
\end{eqnarray}
where 
\begin{equation}
  \label{eq:SandC}
  S_{B_d\to J/\psi
    K_{S,L}}=\frac{2\,\mathrm{Im}(\lambda_{B_d \to J/\psi
    K_{S,L}})}{1+\vert\lambda_{B_d \to J/\psi
    K_{S,L}}\vert^2}\,,\qquad  C_{B_d\to J/\psi
    K_{S,L}}=\frac{1-\vert\lambda_{B_d\to J/\psi
    K_{S,L}}\vert^2}{1+\vert\lambda_{B_d\to J/\psi
    K_{S,L}}\vert^2}\,,
\end{equation}
with
\begin{equation}
  \label{eq:lambda}
  \lambda_{B_d\to J/\psi K_{S,L}}=\eta_{K_S,K_L} (q/p)_{B_d}
\frac{\mathcal{A}(\bar B_d\to J/\psi\,\bar K^0)}{\mathcal{A}(B_d\to J/\psi\,K^0)}(q/p)^*_{K^0}
\end{equation}
and
\begin{equation}
  \label{eq:qop}
  (q/p)_{B_d}=-\frac{V_{tb}^*V_{td}}{V_{tb}V_{td}^*}\,,\qquad
  (q/p)_{K^0}=-\frac{V_{cs}^*V_{cd}}{V_{cs}V_{cd}^*}\,.
\end{equation}
The factors $\eta_{K_S}=-1$ and $\eta_{K_L}=1$ account for the CP eigenvalue
of the final state (neglecting CP violation in kaon mixing).
 
In the limit of vanishing $A_u(B_d,K^0)$, one has
\begin{equation}
\lambda_{B_d\to J/\psi K_{S,L}}=\eta_{K_S,K_L}
\left(
\frac{V_{tb}^*V_{td}}{V_{cb}^*V_{cd}}\right)
\left(\frac{V_{cb}V_{cd}^*}{V_{tb}V_{td}^*}\right)=\eta_{K_S,K_L}e^{-2i\beta}\,,
\label{eq:lambdaSL}
\end{equation}
where the UT angle $\beta=\mathrm{arg}(-(V_{cb}^*V_{cd})/(V_{tb}^*V_{td}))$, so
that
\begin{equation}
  \label{eq:SC0}
  S_{B_d\to J/\psi
    K_{S,L}}=-\eta_{K_S,K_L} \sin 2\beta\,,\qquad C_{B_d\to J/\psi
    K_{S,L}}=0\,.
\end{equation}
A nonvanishing $A_u(B_d,K^0)$ induces a theoretical uncertainy in the
extraction of $\sin 2\beta$ and possibly a value of $C_{B_d\to J/\psi
  K_{S,L}}$ different from zero. Indeed, for $A_u(B_d,K^0)\neq 0$ one
expects a nonvanishing  
\begin{equation}
  \Delta S_{B_d \to J/\psi K_{S,L}}= S_{B_d \to J/\psi K_{S,L}}+
  \eta_{K_{S,L}} \sin 2\beta
  \label{eq:deltaS}
\end{equation}

Let us now discuss how to estimate the value of $A_u(B_d,K^0)$ and
thus the value of $\Delta S_{B_d \to J/\psi K_{S,L}}$ using flavour symmetry.

\section{Evaluation of the theoretical uncertainty of the golden mode
  $\mathbf{B_d \to J/\psi K^0}$ in the Standard Model}
\label{sec:DeltaS}

The basic idea of the $SU(3)$-based methods is to use the flavour symmetry
to extract $A_u(B_d,K^0)$ from a decay channel where $A_u$ is not Cabibbo
suppressed, thereby obtaining an estimate of the departure from
eq.~(\ref{eq:SC0}).
In particular, the method discussed in refs.~\cite{hep-ph/0507290} uses
the two $SU(3)$-related channels $B_d \to J/\psi K^0$ and $B_d \to J/\psi
\pi^0$. The amplitude of $B_d \to J/\psi\pi^0$ can be written as
\begin{eqnarray}
 \mathcal{A}(B_d \to J/\psi\, \pi^0) &=&
  \frac{1}{\sqrt{2}}\left\{\lambda_c^d 
    \left(A_c(B_d,\pi^0)+\Delta_2A_c(B_d,\pi^0)\right)\right.\nonumber\\
 &&- \left.\lambda_u^d \left(A_u(B_d,\pi^0)+\Delta_2A_u(B_d,\pi^0)\right)\right\}\,,
\end{eqnarray}
where, using the notation of ref.~\cite{hep-ph/9812392}, 
\begin{eqnarray}
  A_c(B_d,\pi^0)&=& E_2(c,c,d;B_d,J/\psi,\pi^0) +
  P_2(d,c;B_d,J/\psi,\pi^0)\,,\nonumber \\ 
  \Delta_2A_c(B_d,\pi^0)&=&
  \mathit{EA}_2(c,c,d;B_d,J/\psi,\pi^0)-\mathit{EA}_2(c,c,u;B_d,J/\psi,\pi^0)+\nonumber\\
  &&P_4(d,c;B_d,\pi^0,J/\psi) 
  -P_4(u,c;B_d,\pi^0,J/\psi)\,,\nonumber \\
  A_u(B_d,\pi^0)&=&
  P_2^\mathrm{GIM}(d,c;B_d,J/\psi,\pi^0)-P_2(d,c;B_d,J/\psi,\pi^0)\nonumber\\
  &&-\mathit{EA}_2(u,u,c;B_d,\pi^0,J/\psi)\,,\nonumber  
  \\ 
  \Delta_2 A_u(B_d,\pi^0)&=&
  P_4^\mathrm{GIM}(d,c;B_d,\pi^0,J/\psi)-
  P_4^\mathrm{GIM}(u,c;B_d,\pi^0,J/\psi)-\nonumber\\
  &&P_4(d,c;B_d\pi^0,J/\psi) 
  +P_4(u,c;B_d\pi^0,J/\psi)\,.
\end{eqnarray}
In the $SU(3)$ limit, with the additional assumption of negligible electroweak
penguins ($\Delta_2 A_c$ and $\Delta_2 A_u$) and $\mathit{EA}_2$, one has
$A_c^{SU(3)}=A_c(B_d,K^0)=A_c(B_d,\pi^0)$ and
$A_u^{SU(3)}=A_u(B_d,K^0)=A_u(B_d,\pi^0)$. Therefore there are three
independent hadronic parameters ($\vert A_c^{SU(3)}\vert$, $\vert
A_u^{SU(3)}\vert$ and the relative strong phase) and six measurements
($S$, $C$, and the $CP$-averaged BR in each channel).  Using all these
measurements but $S_{B_d \to J/\psi K_{S,L}}^\mathrm{exp}$, it is possible to
extract the hadronic parameters, thus making a predictions for
$\Delta S_{B_d \to J/\psi K_{S,L}}$. This is the theoretical correction to be used
in the extraction of $\sin2\beta$ from $S_{B_d \to J/\psi K_{S,L}}^\mathrm{exp}$.
For the sake of simplicity, the correlation between $\Delta S_{B_d \to J/\psi K_{S,L}}$
and $S_{B_d \to J/\psi K_{S,L}}^\mathrm{exp}$ is discarded. Its inclusion is
straightforward, but would require the simultaneous fit of the CKM phase and
$\Delta S_{B_d \to J/\psi K_{S,L}}$ within the UT analysis.

Clearly, as $SU(3)$ is not an exact symmetry, the main issue of this
method is to quantify the effect of the $SU(3)$ breaking. In
ref.~\cite{hep-ph/0507290} we tried to reduce the usage of $SU(3)$ to
a minimum, extracting from $B_d \to J/\psi \pi^0$ only the $4\sigma$
range of $\vert A_u^{SU(3)}\vert$, and leaving the phase
unconstrained.  This was a conservative choice in the absence of
independent tests of $SU(3)$ but $A_c(B_d,K^0)\sim
A_c(B_d,\pi^0)$. More recently, a similar estimate of $\Delta S_{B_d
  \to J/\psi K_S}$ using these two decay modes has been presented in
ref.~\cite{arXiv:0809.0842}. The authors of this paper used
exact $SU(3)$ taking
$A_u(B_d,K^0)/A_c(B_d,K^0)=A_u(B_d,\pi^0)/A_c(B_d,\pi^0)$, and
included an estimate of $SU(3)$ breaking in the ratio
$A_c(B_d,K^0)/A_c(B_d,\pi^0)$ based on factorization. In this way they
obtained a negative $\Delta S_{B_d \to J/\psi K_S}$ corresponding to a
shift of $2\beta$ by $[-3.9,-0.8]^\circ$ at $1\sigma$. They also
estimated non-factorizable $SU(3)$-breaking effects, keeping however
the sign of $A_u(B_d,K^0)/A_c(B_d,K^0)$ fixed, obtaining the range
$[-6.7,0]^\circ$.

In the rest of this section we update the analysis of
ref.~\cite{hep-ph/0507290}. In the past five years, the experimental
situation has improved considerably. First, the errors in the two
channels $B_d \to J/\psi K_S$ and $B_d \to J/\psi \pi^0$ have shrunk
by a factor of two. Second, the $BR(B_s \to J/\psi \bar K^0)$ has
been measured, providing an independent test of $SU(3)$.

\begin{table}[tb]
\centering
\begin{tabular}{|cc|cc|}
  $C_1$	 	 & $1.083$               & $C_2$	 & $-0.185$ \\
  $F^{B \to \pi}(m_{J/\psi}^2)$  & $0.4$ & $F^{B \to K}/F^{B \to \pi}$ & $1.2$ \\ 
  $f_{J/\psi}$   & $0.405$               & $m_{B_d}$     & $5.2795$ \\
  $A$            & $0.80 \pm 0.01$       & $\lambda$     & $0.2255 \pm 0.0005$ \\ 
  $\bar\rho $    & $0.164 \pm 0.025$     & $\bar\eta$    & $0.397 \pm 0.023$
\end{tabular}
\caption{Input values used in the analysis. Dimensionful
  quantities are given in GeV.}
\label{tab:inputs}
\end{table}

The input values of the theoretical and CKM parameters used in the present analysis
are given in Table~\ref{tab:inputs}. No error is attached to Wilson coefficients,
form factors and decay constants, as factorized amplitudes only provide the
normalization of the hadronic amplitudes which are fitted from the data.
In particular, we define
\begin{eqnarray}
 A_{u,c}(B_d,\pi^0)&=&\frac{G_F}{\sqrt{2}} m_{B_d}^2 F(B\to\pi) f_{J/\psi}
\left(C_2+\frac{1}{3}C_1\right) {\bar A}_{u,c}(B_d,\pi^0)\,,\nonumber\\
 A_{u,c}(B_d,K)&=&\frac{G_F}{\sqrt{2}} m_{B_d}^2 F(B\to K) f_{J/\psi}
\left(C_2+\frac{1}{3}C_1\right) {\bar A}_{u,c}(B_d,K)\,,
\end{eqnarray}
where $G_F$ is the Fermi constant and the other parameters are listed in 
Table~\ref{tab:inputs}. In the following we give results for the normalized
amplitudes ${\bar A}_{u,c}$. The CKM parameters are taken from the Summer
2010 UT{\em fit} analysis without the $\sin2\beta$ constraint~\cite{utfitsite}.

\begin{table}[tb]
\centering
\begin{tabular}{|cc|cc|}
  $BR^\mathrm{th}$ & $(1.74\pm 0.15)\times 10^{-5}$ &
  $BR^\mathrm{exp}$ & $(1.74\pm 0.15)\times 10^{-5}$\\
  $C^\mathrm{th}$ &
  $ -0.07 \pm 0.11$ &
  $C^\mathrm{exp}$ &
  $ -0.10 \pm 0.13$ \\
  $S^\mathrm{th}$ &
  $-0.95\pm 0.05 \cup -0.82 \pm 0.04$ &
  $S^\mathrm{exp}$ & 
  $-0.93 \pm 0.15$ \\
  $\vert {\bar A}_c\vert$ & $1.21\pm 0.16$ &  & \\
  $\vert {\bar A}_u\vert$ & $0.46\pm 0.46$ &  & \\  
  $\arg{{\bar A}_u}-\arg{{\bar A}_c}$ & $(19\pm 38)^\circ$ & & 
\end{tabular}
\caption{Results of the fit of $\bar B_d \to J/ \psi \pi^0$ (see the
  text for details).}
\label{tab:results2}
\end{table}

\begin{figure}[htb]
\centering
\includegraphics[width=0.3\textwidth]{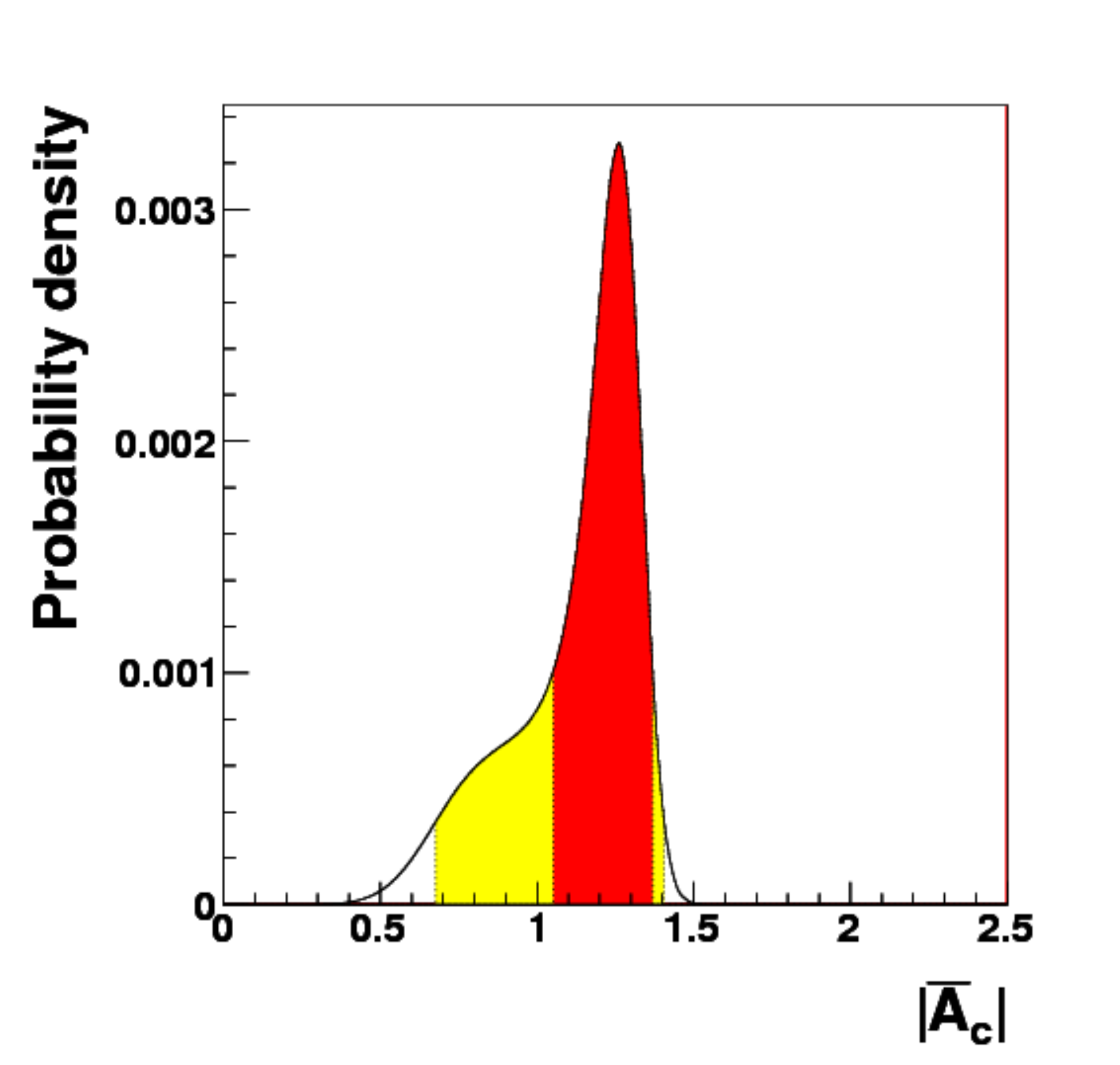}
\includegraphics[width=0.3\textwidth]{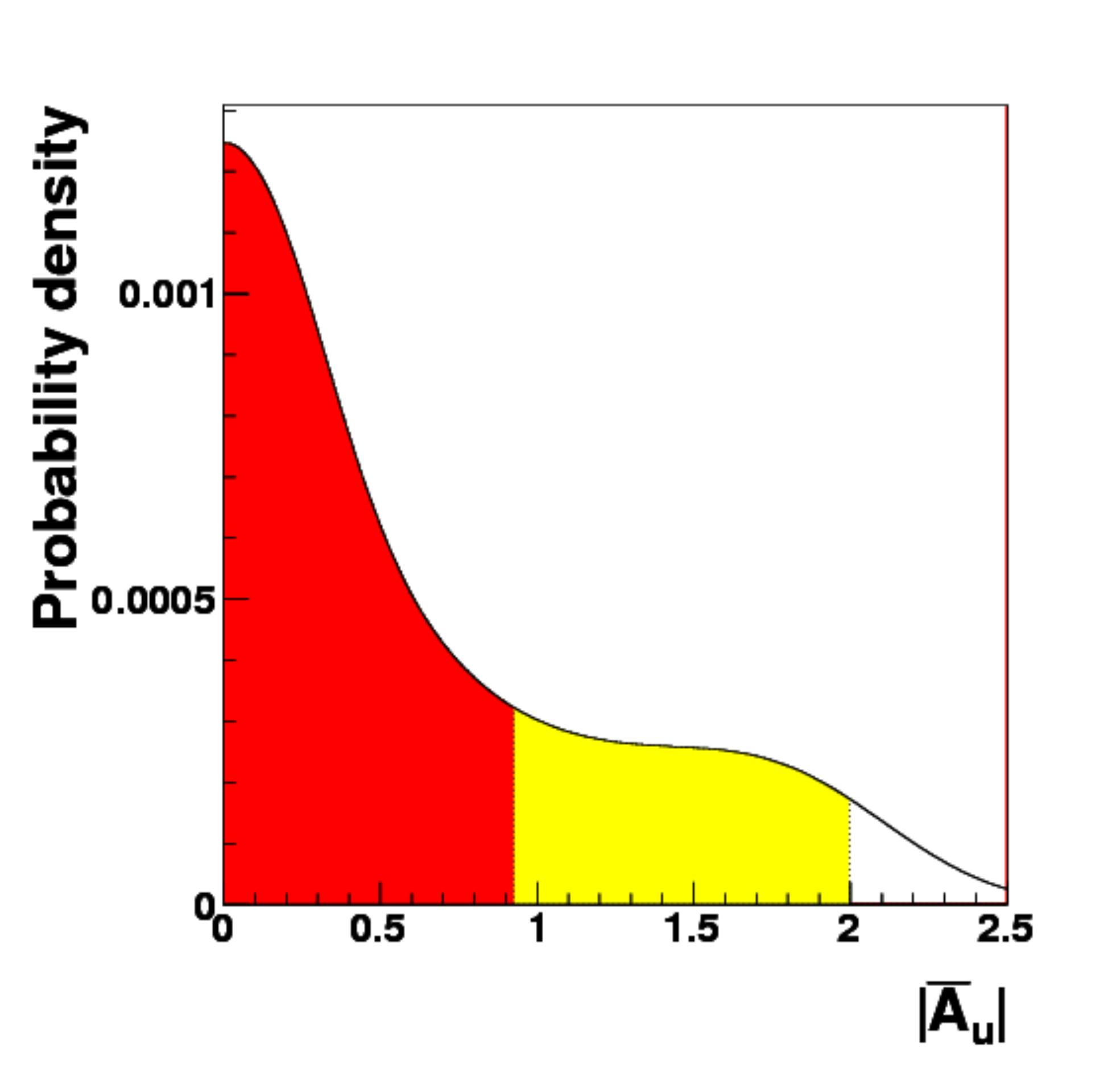}
\includegraphics[width=0.3\textwidth]{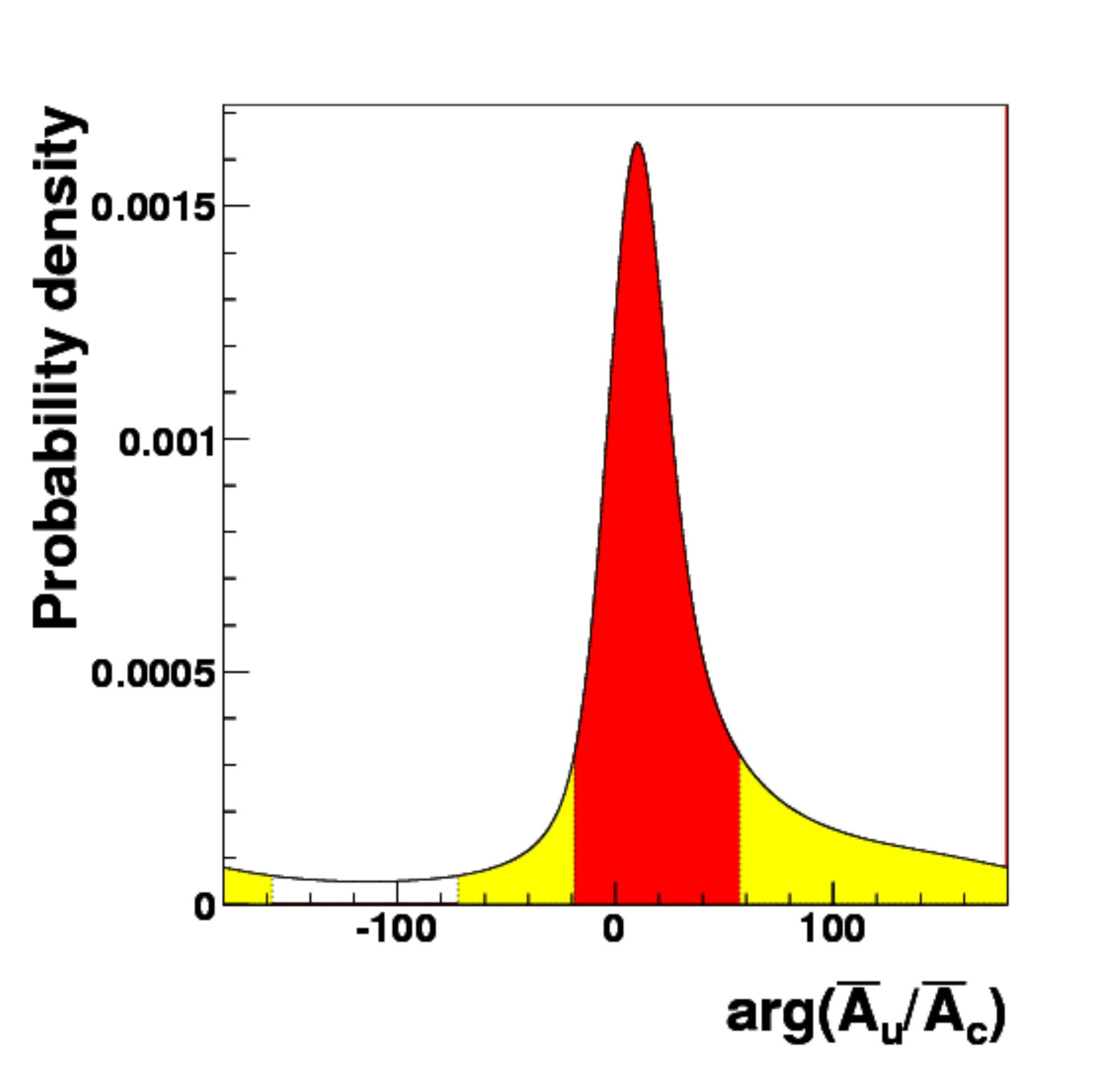}
\caption{Hadronic parameters extracted from $\bar B_d\to J/\psi \pi^0$ data.}
\label{fig:psipi}
\end{figure}

Experimental inputs and results used in the fit of the
$B_d \to J/ \psi \pi^0$ amplitude are given in
Table~\ref{tab:results2}. As also shown in Figure~\ref{fig:psipi},
the value of ${\bar A}_c$ is compatible with one, meaning that, even in the
absence of compelling theoretical arguments, na\"ive factorization provides a
reasonable estimate of this amplitude within $\sim 20$--$30\%$. On the other hand,
$A_u$ is not as well determined. We notice that, with the new data, the
possibity of exchanging the role of $A_c$ and $A_u$ is more disfavoured
than in our previous analysis. Therefore we no longer need to introduce
a cut to retain the $SU(3)$ compatible result only, as we did in
ref.~\cite{hep-ph/0507290}. In addition, the relative strong phase
is now better determined, showing a preference for positive values albeit
with a large uncertainty. In this proceedings, we stick to our original
proposal and discard the phase information (more refined analyses will be
presented in a forthcoming paper). Therefore the $4\sigma$ range
${\bar A}_u < 2.5$ extracted from this fit is the only $SU(3)$-based
information we use to evaluate the theoretical error on $\sin2\beta$.

\begin{figure}[htb]
\centering
\includegraphics[width=0.3\textwidth]{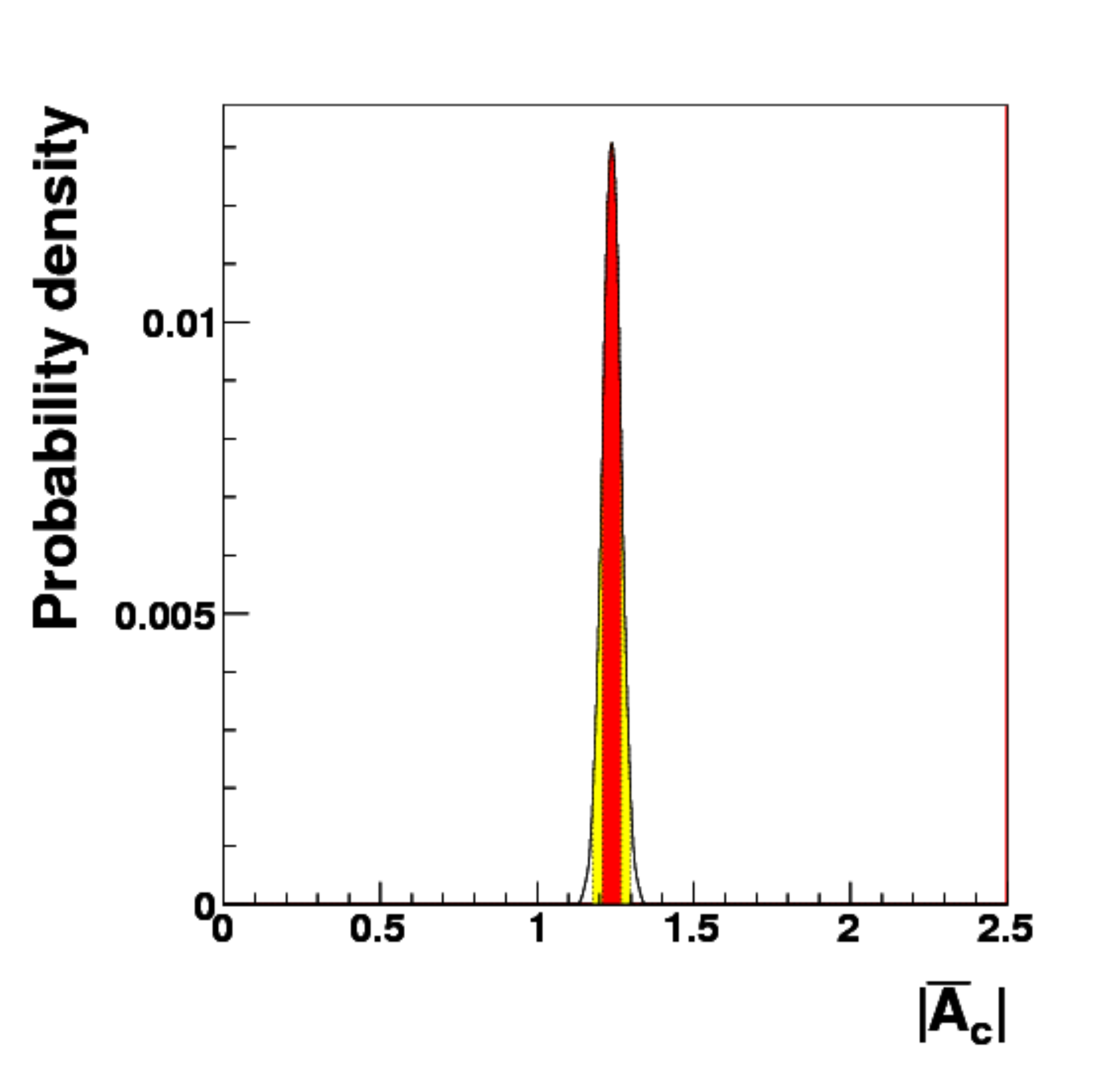}
\includegraphics[width=0.3\textwidth]{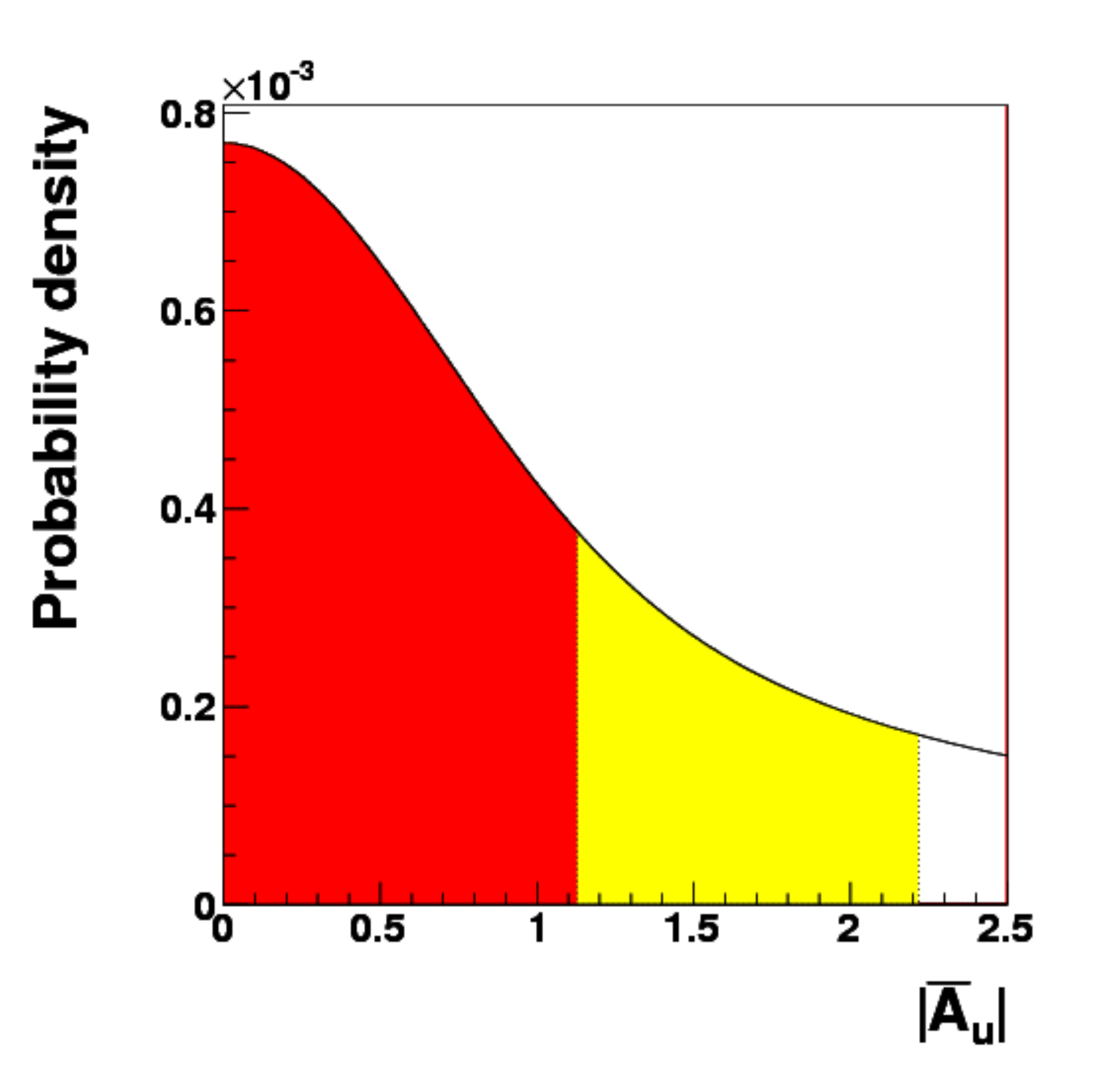}
\includegraphics[width=0.3\textwidth]{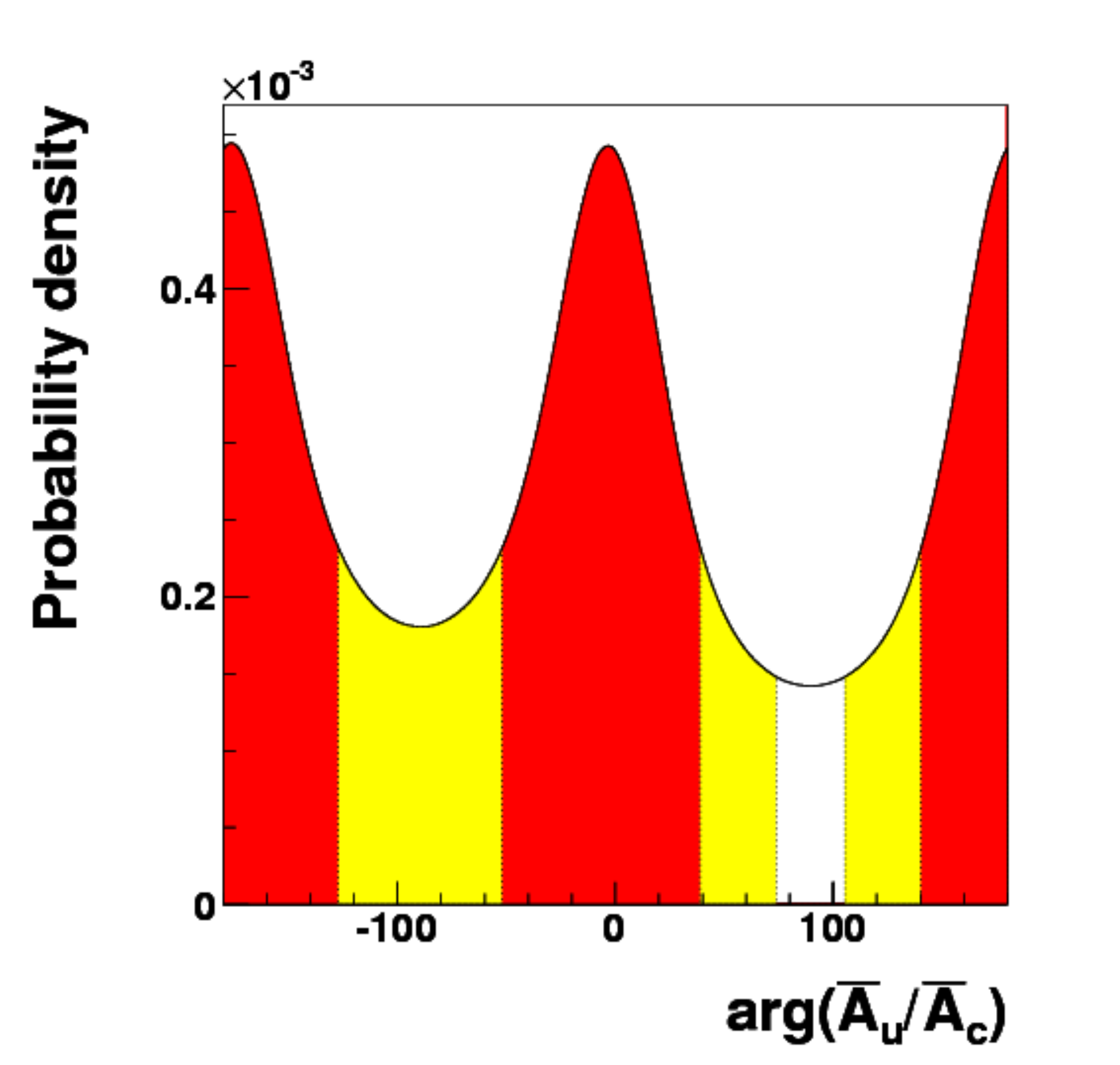}
\caption{Hadronic parameters extracted from $B_d\to J/\psi K^0$ data.}
\label{fig:psiK}
\end{figure}

With this {\em a-priori} cut on $A_u$, we can perform a fit to the $B_d \to J/ \psi K^0$
data. Experimental inputs and results can be found in Table~\ref{tab:results3}. In this
case, $A_c$ is determined much better than in the $B_d \to J/ \psi \pi^0$
case. Again it lies within $\sim 30\%$ of its factorized value, namely it is compatible
with factorization given the typical uncertainties attached to decay constants and form
factors in Table~\ref{tab:inputs}.
As expected, both $A_u$ and the relative phase are practically unconstrained, showing the
importance of the information coming from $B_d \to J/ \psi \pi^0$ for estimating 
$\Delta S$. The corresponfing probability distributions are shown in Figure~\ref{fig:psiK}.

\begin{table}[tb]
\centering
\begin{tabular}{|cc|cc|}
$BR^\mathrm{th}$ & $(8.63\pm 0.35)\times 10^{-4}$ &
$BR^\mathrm{exp}$ & $(8.63\pm 0.35)\times 10^{-4}$ \\
$C^\mathrm{th}$ &
$0.00 \pm 0.01$ &
$C^\mathrm{exp}$ &
$0.00 \pm 0.02$ \\
$S^\mathrm{th}$ &
$0.77 \pm 0.04$ &
$S^\mathrm{exp}$ &
$0.655 \pm 0.024$ \\
$\vert {\bar A}_c\vert $ & 
$1.24 \pm 0.03$ & & \\
$\vert {\bar A}_u\vert$ & $0.56\pm0.56$ & &\\
$\arg{{\bar A}_u}-\arg{{\bar A}_c}$ & $(160\pm 20)^\circ$ & &
\end{tabular}
\caption{Results of the fit of $\bar B_d \to J/ \psi K^0$.
 $S^\mathrm{exp}$ is not used in the fit.}
\label{tab:results3}
\end{table}

Using these results, we find
\begin{equation}
 \Delta S_{B_d \to J/\psi K_S} = 0.00\pm 0.02\,.
\label{eq:deltasfit}
\end{equation}
Since our method discards the phase information on $A_u$ from
$B_d\to J/\psi\pi^0$, the correction we obtain does not shift the
central value of $S_{B_d \to J/\psi K_S}$ but just introduces a
theoretical uncertainty.

Figure~\ref{fig:psiK} shows that the theoretical uncertainty on $\sin2\beta$ is
not entirely negligible with respect to the present experimental error.
We do not find a correction as large as in ref.~\cite{arXiv:0809.0842}, although
the agreement is reasonable considering the aforemetioned differences in the two
methods (notice in addition that the variable  $\Delta\phi_d$ defined in
ref.~\cite{arXiv:0809.0842} to account for the deviation of $S_{B_d \to J/\psi K_S}$
from $\sin2\beta$ is $\Delta\phi_d\sim\Delta S_{B_d \to J/\psi K_S}/\cos2\beta$).

It is very important to stress that the evolution of the $B\to J/\psi\pi$ data is
expected to match the $B\to J/\psi K$ one so that this method will be always able to keep the
theoretical error on the $\sin2\beta$ extraction under control, even reaching the high
precision expected at the super$ B$ factories~\cite{arXiv:0809.0842,O'Leary:2010af}.
LHCb, on the other hand, will be able to exploit the $B_s\to J/\psi K^0$ data to
achieve the same goal with no need of neglecting any hadronic amplitude. 

\section*{Acknowledgements}
M.C. is associated to the Dipartimento di Fisica, Universit\`a di Roma Tre.
L.S. is associated to the Dipartimento di Fisica, Universit\`a di Roma ``La Sapienza''.

\end{document}